
\input phyzzx

\sequentialequations

\overfullrule=0pt
\catcode`\@=11

\def\ns{non-singlet}

\def \F{\phi}

\def \P{\psi}
\def \D{{\delta}}

\def \DM{ {\partial \over {\partial \mu}}}

\def\NP{{\it Nucl. Phys.\ }}

\def\PL{{\it Phys. Lett.\ }}
\def\PR{{\it Phys. Rev.\ }}
\def\PRL{{\it Phys. Rev. Lett.\ }}

\def\IJMP{{\it Int. Jour. Mod. Phys.\ }}
\def\Mod{{\it Mod. Phys. Lett.\ }}

\def\td{two-dimensional}
\def\lc{light-cone}
\def\KT{Kosterlitz-Thouless}
\def\eqaligntwo#1{\null\,\vcenter{\openup\jot\m@th
\ialign{\strut\hfil
$\displaystyle{##}$&$\displaystyle{{}##}$&$\displaystyle{{}##}$\hfil
\crcr#1\crcr}}\,}
\catcode`\@=12

\REF\GM{D.~J.~Gross and A.~A.~Migdal, \PRL {\bf 64} (1990) 717;
M. Douglas and S.~Shenker, \NP {\bf B335} (1990) 635;
E.~Br\'ezin and V.~Kazakov, \PL {\bf 236B} (1990) 144.}
\REF\Doug{ M. Douglas, \PL {\bf 238B} (1990) 176.}
\REF\GMil{D. J. Gross and N. Miljkovi\'c \journal Phys. Lett.
& 238B (1990) 217; E. Br\'ezin, V. Kazakov, and Al. B. Zamolodchikov
\journal Nucl. Phys. &B338 (1990) 673; P. Ginsparg and J. Zinn-Justin
\journal Phys. Lett. &240B (1990) 333; G. Parisi \journal
Phys. Lett. &238B (1990) 209.}
\REF\GKl{D. J. Gross and I. R. Klebanov, \NP {\bf B344} (1990) 475.}
\REF\US{S. Dalley and I. R. Klebanov, Princeton preprint
PUPT-1333, hep-th/9207065.}
\REF\Vort{D. J. Gross and I. R. Klebanov
 \journal Nucl. Phys. &B354 (1991) 459.}
\REF\bk{D. Boulatov and V. Kazakov, Paris preprint LPTENS-91/24.}
\REF\Dall{S. Dalley, \Mod {\bf A7}, 1651 (1992); I. Kostov,
Saclay preprint SPhT/92-096.}
\REF\km{V. Kazakov and A. Migdal, ENS - Princeton preprint LPTENS-92/15;
PUPT-1322.}
\REF\mak{S. Khokhlachev and Yu. Makeenko, Moscow ITEP preprint ITEP-YM-7-92.}
\REF\qcd{G. 't Hooft, \NP {\bf B75} (1974) 461.}
\REF\Thorn{C. Thorn, \PL {\bf 70B} (1977) 85; \PR {\bf D17} (1978)
1073.}
\REF\Brod{H.-C. Pauli and S. Brodsky, \PR {\bf D32} (1985) 1993, 2001;
K. Hornbostel, S. Brodsky and H.-C. Pauli, \PR {\bf D41} (1990) 3814;
for a good review, see K. Hornbostel, Ph. D. thesis, SLAC report
No. 333 (1988).}
\REF\BPR{W. A. Bardeen, R. B. Pearson and E. Rabinovici,
\PR {\bf D21} (1980) 1037.}
\REF\KS{I. R. Klebanov and L. Susskind, \NP {\bf B309}, 175 (1988).}
\REF\DM{S. Dalley and T.R. Morris, \IJMP {\bf A5}, 3929 (1990).}
\REF\mar{E. Marinari and G. Parisi, \PL {\bf B240} (1990) 375.}
%
%
%
%
%

\def\eqaligntwo#1{\null\,\vcenter{\openup\jot\m@th
\ialign{\strut\hfil
$\displaystyle{##}$&$\displaystyle{{}##}$&$\displaystyle{{}##}$\hfil
\crcr#1\crcr}}\,}
\catcode`\@=12

\def\oh{{1 \over 2}}
\def\b{\beta}
\def\a{\alpha}

\def\half{{1\over 2}}
\def\d{\dagger}
\def\tj{{\tilde{J}^{+}}}
\def\pa{\partial}

\nopagenumbers

{\baselineskip=12pt
\line{\hfil PUPT-1342}
\line{\hfil hep-th/9209049}
}
\title{String Spectrum of 1+1-Dimensional Large $N$
QCD with Adjoint Matter}
\author{Simon Dalley and Igor R. Klebanov }
\JHL
\abstract
We propose gauging matrix models of string theory to eliminate unwanted
non-singlet states. To this end
we perform a discretised \lc\ quantisation of large $N$ gauge theory
in 1+1 dimensions, with scalar or fermionic matter fields transforming
in the adjoint
representation of $SU(N)$.
The entire spectrum  consists of bosonic and fermionic
closed-string excitations,
which are free as $N \to \infty$. We analyze the  general features of
such  states as a function of the cut-off
and the gauge coupling, obtaining  good
convergence for the case of adjoint fermions. We discuss possible
extensions of the model and the search for new non-critical string theories.
\endpage

\pagenumbers
\vsize=8.9in
\hsize=6.5in
\centerline{\caps 1. Introduction}
\bigskip

The construction of string theories in non-critical dimensions is
an interesting problem which remains largely unsolved.
A fruitful idea is to study large-$N$ field theories, where
the Feynman diagrams are grouped according to their two-dimensional
topology and, after some adjustment of parameters, can be identified
with string worldsheets. Recently this idea has been extensively
developed to sum over all worldsheets embedded in dimensions $c\leq 1$
[\GM-\GKl].
It is of great interest to extend this progress to higher-dimensional
embeddings.

In a previous paper [\US] we considered a two-dimensional interacting
scalar matrix field theory with a global $SU(N)$ symmetry, where the Feynman
graphs (without tadpoles) are finite. In such a theory the sum over
the planar graphs is expected to possess a critical point similar
to those found for $c\leq 1$. At this critical point the model should
describe two-dimensional quantum gravity coupled to $c=2$ matter.
Our \lc\ analysis indeed suggested that the critical point exists.
The large-$N$ scalar field theory is a straightforward extension
to higher dimension of the technique that has been so successful for
$c\leq 1$. Is this a viable candidate for non-critical string theory in
any dimension? We believe that the answer is negative. The main
reason is the lack of decoupling of the $SU(N)$ \ns\ states, which
have infinite degeneracies and cannot be regarded as closed-string states.
The \ns s are like strings that have disintegrated into separate
bits. It is not hard to show [\US] that the \ns s have finite energy
and, therefore, do not decouple. This should be contrasted with the $c=1$
case, where the \ns s are pushed to infinite energy in the continuum
limit [\GKl, \Vort, \bk].
The basic physical reason for this difference is perhaps that,
in the $0+1$-dimensional theory ($c=1$) the space consists of
only a point and a string cannot fall apart into its elementary
constituents. In $1+1$ dimensions, however, the question of string
stability already becomes crucial. For a large-$N$ model to be a good
string theory, it is necessary that the field quanta (string bits)
are absolutely bound into $SU(N)$ invariant states (the strings).
In a scalar theory we may find some attraction of the quanta, as
suggested by our results in [\US], but the binding is not absolute.
Therefore at sufficiently high energies the strings fall apart
and the model does not make sense as a theory of elementary string
excitations.

In [\US] we suggested an obvious cure for this problem: introduction of
a gauge field. The resulting confinement pushes the \ns\ states to
infinite energy, so that the string bits become absolutely bound into
strings. Let us note that all the $c\leq 1$ models can be reformulated
as gauge theories [\Dall]. For instance, the gauged matrix quantum
mechanics ($c=1$) has the lagrangian
$$L=\Tr \left\{ \half (\dot\phi+i[A, \phi])^2-V(\phi)\right\}
\ .\eqn\eq$$
The corresponding hamiltonian is
$$ H=
\Tr \left\{ \half \Pi^2+V(\phi)+i[\Pi, \phi] A\right\}
\ .\eqn\eq$$
Thus, $A$ is simply a Lagrange multiplier which
imposes the condition $i[\Pi, \phi]|\Psi>=0$ on the physical
states. Only the $SU(N)$ singlet wave functions satisfy this constraint,
and the non-singlets are discarded.
In the ungauged models the energies of the non-singlets diverge at the critical
point of continuous worldsheets [\GKl, \Vort, \bk], resulting
in equivalence with the gauged models at this point.\foot{The case
of compact $c=1$
is somewhat more subtle because there the non-singlets implement the effects of
\KT\ vortices [\GKl, \Vort, \bk, \Dall].}

In this paper we consider the gauged matrix models in 1+1 dimensions,
{\it i.e.} $SU(N)$ gauge theories coupled to matter fields
in the adjoint representation\foot{The infinite coupling limit
of such models has recently been studied on a lattice with a rather
different motivation [\km, \mak].}. Here the effect of gauge fields
is absolutely crucial because they produce confinement which
otherwise would not exist.
Also the two-dimensional nature of the problem
means that there are no physical gluons and so strings are still built from
the matter quanta alone. We carry out \lc\ quantization of these
models as $N \to \infty$ and study them as a function of
the gauge coupling strength, this being of some interest in itself apart from
the string interpretation.
Our approach is similar to `t Hooft's classic solution [\qcd] of the
\td\ large-$N$ gauge theory coupled to fermions in the fundamental
representation. In that case quark and anti-quark pairs are confined into
mesons whose masses are found from a simple bound state integral
equation. This theory can be thought of as an open string model,
the spectrum forming a single rising ``Regge trajectory''. We will
argue that the coupling to adjoint matter fields produces a kind of
closed string analogue of the `t Hooft model. This
is much more complicated than `t Hooft's open string because our closed
string can contain any number of matter quanta (string bits),
the \lc\ hamiltonian
connecting different number sectors. Therefore the
bound state equation is hard to write down explicitly.
However, if we introduce a cut-off in the form of discretized
longitudinal momentum [\Thorn, \Brod],
the bound state equation can be set up easily
and reduces to a matrix eigenvalue problem, with the size of the matrix
diverging as the cut-off is removed. Even for relatively low
values of the cut-off, some qualitative and sometimes quantitative
conclusions can be drawn from this method.

The adjoint matter field can be taken to be either a scalar or a
fermion, and we analyze both cases. The scalar case gives a kind of
closed bosonic string theory with rising ``Regge trajectories''.
The adjoint fermion theory is perhaps more intriguing. It can be
interpreted as a fermionic string whose spectrum contains both bosons
(strings with even numbers of bits) and fermions (strings with odd
numbers of bits). The discretized free string eigenvalue equation
can be easily set up for both types of states. Low-lying states are found
to be spectacularly pure in the sense that the mass eigenstates are almost
exactly eigenstates of string length (number of matter quanta). This
encourages us to go to much higher cutoffs by diagonalising in the subspace of
states of a given length.

In section 2 we introduce the models and carry out \lc\ quantization.
In section
3 we introduce the discretization of longitudinal momentum and reduce the free
string theory to a matrix eigenvalue problem. We give some of our
numerical results on the spectrum as a function of the gauge coupling and
cutoff.
In section 4 we discuss possible extensions of this
work to higher dimensions, supersymmetric spectrum, etc.. and speculate
on the search for new non-critical string theories.
\bigskip
\centerline{\caps 2. Light-cone quantisation}
\bigskip

In this section we consider the light-cone quantization of
$1+1$-dimensional
$SU(N)$ gauge theory coupled to matter in the adjoint representation.
Strings will arise as collective excitations and, depending on whether
the matter field is a scalar or a fermion, will be of bosonic or bosonic and
fermionic type.

\undertext{Scalar Matter}

Here the Minkowski space action is taken to be
$$ S_{sc} = \int d^2 x \Tr \left[ \oh D_{\a}\F D^{\a}\F -\oh m^2 {\F}^2
-{1\over 4 g^{2}} F_{\a \b}F^{\a \b}\right], \eqn\action$$
where $F_{\a \b} = \pa_{\a} A_{\b} -\pa_{\b}
A_{\a} + i[A_{\a},A_{\b}]$ and the covariant derivative is
$D_{\a}\phi = \pa_{\a}\phi
+i[A_{\a}, \phi]$.
The scalar $\F$ and the gauge potential $A_\alpha$ are
traceless $N$x$N$ hermitian matrix fields.
To leading order in $N$ it is not necessary to
impose the tracelessness condition from the beginning. Instead, as we
will show later, this condition is implemented by a simple restriction
on the physical states.
A similar theory, with $\phi$ taken as a general complex matrix,
 was considered by Bardeen, Pearson and Rabinovici [\BPR]
as part of their attempt to solve higher-dimensional pure gauge theory.
We have not included explicit self-interactions of the matter
fields in \action\ but we shall return to this question in section 4 where
we discuss the possibility of finding new critical behaviour.

An interesting feature of the theory in eq. \action\ is that it
corresponds to dimensional reduction of the 2+1-dimensional
pure gauge theory. If the 3-d gauge field $A_\mu=(A_\alpha, A_3)$
is taken to be independent of $x^3$, then the action reduces to
eq. \action\ with $m=0$ and $\phi\sim A_3$. Such a reduction is valid
when $x^3$ is compact and small, for example at high temperature in the
euclidean version. In fact polynomials of $\phi$ can also be obtained in the
reduced action by adding appropriate combinations of Wilson lines winding
around the $x^3$ direction. The reduction generalises to higher
numbers of compact and non-compact dimensions also,
involving gauged multi-matrix models etc..

To perform \lc\ quantisation of  \action\
we introduce variables $x^{\pm}=(x^0\pm x^1)/\sqrt 2$ and treat $x^{+}$ as
the time variable. Our conventions will be such that $g^{+-}=g^{-+}=1$ so
for any 2-vector $k^{\alpha}$
we have $k^{+}=k_{-}$. As is standard we choose the
convenient light-cone gauge $A_{-}=0$; there are no Fadeev-Popov ghosts, and
no need for a gauge fixing term since there are no dynamical degrees of
freedom from the gauge potential to be quantized. The action reduces to
$$S_{sc}=\int dx^{+}dx^{-} \Tr \left[\pa_{+}\F\pa_{-}\F - \half m^2 \F^2
+{1\over 2 g^2} (\pa_{-}A_{+})^2 + A_{+}J^{+} \right] \eqn\mact$$
where the longitudinal momentum current $J^{+}_{ij} = i[\F,\pa_{-}\F]_{ij} $.
 The action does not depend on
$x^{+}$ `time' derviatives of the gauge potential $A_{+}$.
Therefore $A_{+}$ is not dynamical and may be eliminated
by the constraint of its Euler-Lagrange equations. It is convenient
to split the gauge field into its zero and non-zero modes as
$A_+=A_{+,0}+{\bar A}_+$. Then the constraints are
$$\eqalign{ \int dx^- J^+ =0\ ,& \cr
\pa_{-}^{2} {\bar A}_{+} - g^2 J^{+} & = 0\ . \cr
}\eqn\const$$
The first of these equations is enforced by $A_{+, 0}$ which
acts as a lagrange multiplier.

We can now proceed to quantise the
remaining matter degrees of freedom.
The \lc\ momentum and
energy  $P^{\pm}= \int dx^- T^{+\pm}$ are found to be
$$ \eqalign{&P^{+} =  \int dx^{-} \Tr
[(\partial_{-}\F)^{2} ]\ ,\cr
&P^{-} = \int dx^{-} \Tr \left[\half m^2 \F^2
-  \half g^2  J^{+} {1\over \pa_{-}^{2}} J^{+}
\right]\ .\cr }\eqn\pminus$$
Given an initial field configuration on the line $x^{+}=0$ we obtain the
field at later times by Hamiltonian developement.
We choose the free field representation
 for $\F(x^+=0)$, with mode
decomposition\foot{The symbol $\dagger$ is from here on always understood
as the  quantum version of complex conjugation and does not act on indices.};
$$\F_{ij}={1 \over \sqrt{2\pi}} \int_{0}^{\infty} {dk^{+}
\over \sqrt{2k^{+}}} \left(a_{ij}(k^{+}){\rm e}^{-ik^{+}x^{-}} +
a_{ji}^{\d}(k^{+}) {\rm e}^{ik^{+}x^{-}}\right)\ .
\eqn\Four$$
As usual in \lc\ quantisation, positive energy quanta have positive \lc\
momentum $k^+$; this fact leads in large part to the successful employment
of the discretised approach we formulate later.
The modes $a_{ij}$ are normalised so that when we impose the canonical
commutation
relation at $x^{+}=0$;
$$[\F_{ij}(x^{-}),\pa_{-}\F_{kl}(\tilde{x}^{-})] =
\half i
\delta(x^{-}-\tilde{x}^{-})\delta_{il} \delta_{jk}\ ,\eqn\ccr$$
we obtain
$$ [a_{ij}(k^{+}),a_{lk}^{\d}(\tilde{k}^{+})] =
\delta(k^{+} - \tilde{k}^{+})
\delta_{il} \delta_{jk} \ .\eqn\modeccr$$
Substituting the mode expansion \Four\ into eq. \pminus\ and normal ordering,
we can derive explicit oscillator expressions. Thus, we find
$$P^+ = \int_{0}^{\infty} dk\ k a_{ij}^{\d}(k)a_{ij}(k)\ \eqn\nomess$$
The hamiltonian $P^-$ of the $m \to 0$ (strong coupling) limit of the
theory can be neatly written as
(repeated indices are summed over);
$$ P^-(m \to 0)= g^2\int_0^\infty
 {dk^+\over (k^+)^2} \tilde J^+_{ik} (k^+) \tilde J^+_{ki} (-k^+)
\eqn\strong$$
where we introduced the current Fourier transform
$$\tj (k^+)= {1\over \sqrt{2\pi}}
\int dx^- J^{+}(x^-) {\rm exp}(-ik^+x^-)\ .$$
We will often drop the superscript on $k^+$ for brevity.
Explicitly we find, for $q>0$,
$$\eqalign{2\sqrt{2\pi}\tj_{ki} (-q) = & \int_{0}^{\infty} dp {2p+q \over
\sqrt{p(p+q)}}
\bigl(a^{\d}_{jk}(p)a_{ji}(p+q) - a^{\d}_{ij}(p)a_{kj}(p+q)\bigr) \cr
& + \int_{0}^{q} dp {q-2p \over \sqrt{p(q-p)}}a_{kj}(p)a_{ji}(q-p)
\cr }\eqn\jay$$
and $\tj_{ik} (q)=(\tj_{ki} (-q))^\dagger$.
Implicit in eq. \strong\ is a choice
of normal ordering that preserves positivity and ensures that
the light-cone oscillator vacuum $|0>$ is the ground state of $P^-$
with eigenvalue zero.

In general, a basis for the Fock space states can be taken
of the form;
$$a_{i_{1}j_{1}}^{\d}(k_{1}^{+})\cdots a_{i_{b}j_{b}}^{\d}(k_{b}^{+})
|0>\ ;\qquad\qquad a_{ij}(k^{+})|0>=0 \ .\eqn\Fock$$
In a gauge theory however, there is a dramatic reduction in
the number of physical states due to the gauge invariance.
Namely, the quantum version of the first constraint in eq.
\const\ acts as the condition on physical states
$$\tj_{ki} (q=0)|\Psi> = 0\ .
\eqn\scons$$
It follows that the $k^+$-integral in eq. \strong\ has to be defined
in a principal value sense, and that the energies of physical states
stay finite.
It is not hard to check that the only states that satisfy condition
\scons\ are the singlets under the residual global $SU(N)$ symmetry;
$$ N^{-b/2} \Tr [ a^{\d}(k_{1}^{+}) \cdots a^{\d}(k_{b}^{+})] |0> \eqn\state$$
As we discussed in [\US], these are precisely the states that have a
nice closed string interpretation. For instance, \state\ can be
thought of as
a single closed string
consisting of $b$ bits, each bit carrying a proportion of the total
longitudinal momentum of the string $P^+ = \sum_{i=1}^{b}k^{+}_{i} $.
To leading order in $N$, the only effect of the condition
$\Tr \phi=0$ on our quantization procedure is to exclude the one-bit state
$\Tr [a^\dagger (P^+)]|0>$.

For comparison, recall that in the scalar
theory with self-interaction but no gauge fields it was found in [\US] that the
non-singlet states, which have no simple closed string interpretation, were
always of finite energy. After gauging the theory,
the trivial confinement in
two dimensions pushes these states to infinite energy.
This apparently removes the main obstacle to building a
connection between string theories and large-$N$ models in dimensions
greater than one.

In the limit $N \to \infty$ the \lc\ hamiltonian $P^-$ propogates single closed
string states to single closed string states. Terms that convert one closed
string into two closed strings (two oscillator traces acting
on the vacuum) are of
order $1/N$. Thus, as expected, the string coupling constant is $\sim 1/N$
and sending it to zero will allow us to study the spectrum of free closed
strings. Practically speaking this limit means that the hamiltonian
$P^-$ acts locally on the string, splitting bits  and joining
adjacent bits in the trace. Explicitly, we find;
$$\eqalign{&P^{-} =  \half m^2\int_{0}^{\infty} {dk\over k} a_{ij}^{\d}(k)
a_{ij}(k) +{g^2 N\over 4\pi}\int_{0}^{\infty} {dk\over k}\
C a_{ij}^{\d}(k)a_{ij}(k) \cr
&+ {g^2\over 8\pi}
 \int_{0}^{\infty} {dk_{1} dk_{2} dk_{3} dk_{4} \over
\sqrt{k_{1}k_{2}k_{3}k_{4}}} \biggl\{ A \D (k_{1}+k_{2}-k_{3}-k_{4})
a_{kj}^{\d}(k_{3})a_{ji}^{\d}(k_{4})a_{kl}(k_{1})a_{li}(k_{2})  \cr
& + B \D(k_{1} + k_{2} +k_{3} -k_{4})
\bigl(a_{kj}^{\d}(k_{1})a_{jl}^{\d}(k_{2})
a_{li}^{\d}(k_{3})a_{ki}(k_{4}) + a_{kj}^{\d}(k_{4})a_{kl}(k_{1})a_{li}(k_{2})
a_{ij}(k_{3})\bigr) \biggr\} \cr }\ .\eqn\mess$$
The coefficients $A,B$ and $C$ are;
$$\eqalign{& A=
 {(k_{2}-k_{1})(k_{4}-k_{3}) \over (k_{1} + k_{2})^2 } -
{(k_{3} + k_{1})(k_{4} + k_{2}) \over (k_{4}-k_{2})^2 }  \ ,\cr
&B=  {(k_{1}+k_{4})(k_{3}-k_{2}) \over (k_{3}+k_{2})^2 } +
{(k_{3}+k_{4})(k_{1}-k_{2}) \over (k_{1}+k_{2})^2 }  \ ,\cr
&C= \int_{0}^{k} dp {(k+p)^2 \over p(k-p)^2 } \ .\cr }\eqn\coefbos$$
As we argued previously,
those momentum integrals in \mess\ which have singularities are
understood in the principal value sense. We shall not specify this more
precisely since in the following section we will discretise the momentum
integrals explicitly for numerical computation, in which case the finiteness
of the spectrum will be easy to demonstrate. Also the `self-induced
inertia'  $C$ has been kept explicit and not absorbed in a renormalisation of
the mass since it really removes a divergence present in the quartic
terms.  Though the dynamics governed by $P^-$ are quite complicated, each
term in \mess\ has a simple interpretation. The mass term represents
 a potential
which measures the tensional energy of a string without changing its state.
The kinetic terms either take three neighboring bits into one or one bit into
three, or redistribute the momentum in two adjacent bits.
(Obviously this does not mix states with odd and even numbers of
bits, and these two sectors can be treated separately).
These dynamics are
to be understood by analogy with the fluctuations of the Liouville degree of
freedom in non-critical string theory. The \lc\ formalism for the
$c=2$ case, which leads to a string hamiltonian similar to \mess,
was developed in [\US]. However, as we explained, there are significant
physical differences between the pure scalar model and the gauge theory
which manifest themselves in the absence of the non-singlet states in
the latter.
We will also find that the gauged model has no phase transition as
a function of the only dimensionless parameter $g/m^2$. Perhaps, as
we suggest in section 4,
critical behaviour can be found in a gauged model with the quartic
scalar potential.

\bigskip
\undertext{Fermionic Matter}

In this case the action is taken to be
$$ S_f = \int d^2 x \Tr \left[
i\Psi^T\gamma^{0} \gamma^{\a} D_{\a} \Psi -m\Psi^{T}\gamma^{0} \Psi
-{1\over 4 g^{2}} F_{\a \b}F^{\a \b}\right] \eqn\faction$$
where $\Psi_{ij}=2^{-1/4} {\psi_{ij}\choose \chi_{ij}}$
is a two-component spinor,
and the transposition acts only on the Dirac indices,
$\Psi^T_{ij}=2^{-1/4} (\psi_{ij}\ \chi_{ij})$.
$\chi$ and $\P$ are  traceless hermitian
$N$x$N$ matrices of grassmann variables,
$$\chi_{ij}^\star=\chi_{ji}\ , \qquad\qquad
\psi_{ij}^\star=\psi_{ji}\ ,
$$
and the covariant derivative is defined by
$D_{\a}\Psi = \pa_{\a}\Psi
+i[A_{\a}, \Psi]$. Choosing the light-cone gauge $A_-=0$, and
the `chiral' representation $\gamma^0=\sigma_2$, $\gamma^1=i\sigma_1$,
we find the action
$$S_f=\int dx^{+}dx^{-} \Tr \left[
i\psi \pa_{+} \psi + i\chi \pa_{-} \chi -i\sqrt 2 m \chi \psi
+{1\over 2 g^2} (\pa_{-}A_{+})^2 + A_{+}J^{+} \right] \eqn\mact$$
where the longitudinal momentum current is now of the form
$J^{+}_{ij} = 2\psi_{ik} \psi_{kj}$.
The gauge potential $A_+$ and the
left-moving fermion $\chi$ are non-dynamical
degrees of freedom and can be eliminated by their constraint equations.
Two of the constraints are identical to those in the scalar case,
eq. \const, while $\chi$ is determined by
$$
\sqrt{2} \pa_{-}\chi - m\psi   = 0\ . \eqn\fermcon
$$
Using the constraints, we find that eq. \pminus\ of the scalar case
is replaced by
$$ \eqalign{&P^{+} =  \int dx^{-} \Tr
[i\psi \pa_{-} \psi]\ ,\cr
&P^{-} = \int dx^{-} \Tr \left[-{im^2 \over 2}
\psi {1\over \pa_{-}} \psi -  \half g^2  J^{+} {1\over \pa_{-}^{2}} J^{+}
\right]\ .\cr }\eqn\fminus$$
Now we introduce the mode expansion,
$$\psi_{ij} = {1\over 2\sqrt\pi} \int_{0}^{\infty} dk^{+}
\left(b_{ij}(k^{+}){\rm e}^{-ik^{+}x^{-}} +
b_{ji}^{\d}(k^{+}){\rm e}^{ik^{+}x^{-}}\right )\ .\eqn\mode $$
 From the canonical commutation relations
$$\{\psi_{ij}(x^{-}), \psi_{kl}(\tilde{x}^{-})\} =
\half
\delta(x^{-}-\tilde{x}^{-})\delta_{il} \delta_{jk}\eqn\anticomm$$
it follows that
$$\{b_{ij}(k^{+}), b_{lk}^{\d}(\tilde{k}^{+})\} =
\delta(k^{+} - \tilde{k}^{+})
\delta_{il} \delta_{jk} \eqn\modac$$
In terms of the oscillators, eq. \fminus\ assumes the form
$$P^+ = \int_{0}^{\infty} dk\ k b_{ij}^{\d}(k)b_{ij}(k)\ ,\eqn\fp$$
$$\eqalign{&P^{-} =  \half m^2\int_{0}^{\infty} {dk\over k} b_{ij}^{\d}(k)
b_{ij}(k) +{g^2 N\over \pi} \int_{0}^{\infty} {dk\over k}\
C b_{ij}^{\d}(k)b_{ij}(k) \cr
&+ {g^2\over 2\pi} \int_{0}^{\infty} dk_{1} dk_{2} dk_{3} dk_{4}
\biggl\{ A \D (k_{1}+k_{2}-k_{3}-k_{4})
b_{kj}^{\d}(k_{3})b_{ji}^{\d}(k_{4})b_{kl}(k_{1})b_{li}(k_{2})  \cr
& + B \D(k_{1} + k_{2} +k_{3} -k_{4})
(b_{kj}^{\d}(k_{4})b_{kl}(k_{1})b_{li}(k_{2})
b_{ij}(k_{3})-
b_{kj}^{\d}(k_{1})b_{jl}^{\d}(k_{2})
b_{li}^{\d}(k_{3})b_{ki}(k_{4})) \biggl\} \cr }\eqn\fpminus$$
where the coefficients are now given by
$$\eqalign{&A= {1\over (k_{4}-k_{2})^2 } - {1\over (k_{1}+k_{2})^2}\ , \cr
&B= {1\over (k_{2}+k_{3})^2 } - {1\over (k_{1}+k_{2})^2 }\ , \cr
&C= \int_{0}^{k} dp {k\over (p-k)^2}\ . \cr }\eqn\coeffer$$

The discussion of the space of states is similar to that
in the scalar case. Again, the confinement requires the
physical states to be $SU(N)$ singlets. One important new feature
is that the strings with odd numbers of bits are fermions, while
the strings with even numbers of bits are bosons. To
leading order in $1/N$ (free string approximation) these two sectors
do not mix. They will be discussed separately in section 3.
\bigskip
\centerline{\caps 3. Discretized formulation and numerical results}
\bigskip

In order to study the theories of section 2, we will introduce a
cut-off in the form of discretized longitudinal momentum [\Thorn, \Brod].
Now $k^+$ is only allowed to take values
$nP^{+}/K$ where the integer $n \leq K $ and $K \to \infty$ in the
continuum limit.
This can be arranged by making $x^-$ compact and
adopting periodic boundary conditions on the matter fields [\Brod].
Moreover in a massive theory quanta with $k^+=0$ are excluded because
they carry infinite energy.
Therefore $n$ is restricted to be a positive
integer and the mode expansions become
$$\eqalign{
&\phi_{ij} ={1 \over \sqrt{4\pi}} \sum_{n=1}^\infty {1\over \sqrt{n}}
\left (A_{ij}(n){\rm e}^{-iP^{+}nx^-/K} + A_{ji}^{\d}(n)
{\rm e}^{iP^{+}nx^-/K} \right )\ ,\cr
&\psi_{ij} ={1 \over \sqrt{4\pi}} \sum_{n=1}^\infty
\left (B_{ij}(n){\rm e}^{-iP^{+}nx^-/K} + B_{ji}^{\d}(n)
{\rm e}^{iP^{+}nx^-/K} \right )\ ,\cr }
\eqn\dm$$
with the oscillator algebra
$$[A_{ij}(n),A_{lk}^{\d}(n')] =
\{ B_{ij}(n),B_{lk}^{\d}(n')\} =
\delta_{n n'}
\delta_{il}\delta_{jk}\ .\eqn\modecomm$$

In the case of the scalar matter,
a normalized state of $b$ bits is of the form
$${1\over N^{b/2}\sqrt{s}}
\Tr [A^{\d}(n_1)\cdots A^{\d}(n_b)] |0>\ .\eqn\dstring$$
The states are defined by
ordered partitions of
$K$ into $b$ positive integers, modulo cyclic permutations.
Therefore the closed strings are oriented.
If $(n_1,~ n_2,~ \ldots,~ n_b)$ is taken into itself
by $s$ out of $b$ possible cyclic permutations,
then the corresponding state receives a
normalization factor $1/\sqrt{s}$. In the absence of special
symmetries, $s=1$.

For the fermionic matter, if $b$ is odd (fermionic states)
the discussion above goes through after replacing $A_{ij}$
by $B_{ij}$. Thus the normalized states
$${1\over N^{b/2}\sqrt{s}}
\Tr [B^{\d}(n_1)\cdots B^{\d}(n_b)] |0> \eqn\fstring$$
are again in one-to-one correspondence with
ordered partitions of
$K$ into $b$ positive integers, modulo cyclic permutations.
For even $b$ however, this is no longer the case since
some states vanish due to the fermionic statistics of the oscillators.
A simple example is
$$\Tr [B^{\d}(n_1) B^{\d}(n_1)] |0>=0\ .\eqn\vstring$$
In general, all partitions of $K$ where $b/s$ is odd do not
give rise to states. For even $b/s$, the normalized states are constructed
in the same way as before and are given by eq. \fstring.

In all cases the restriction $\sum_{i=1}^b n_i=K$ on the positive
integers $n_i$ makes the total number of states finite. This
allows for an explicit diagonalization of the (mass)$^2$ operator
$2P^+P^-$. The goal is to study the mass spectrum
as a function of the increasing cut-off $K$, looking for convergence
of the low-lying masses. For the scalar matter, we obtain
$$2P^+P^{-} = {g^2 N\over 4\pi}~K\left(y V_{sc} +T_{sc}\right)
\ ,\eqn\ham$$
where
$$V_{sc}=\sum_{n=1}^{\infty} {1\over n}
A^{\d}_{ij}(n)A_{ij}(n) \eqn\pot$$
is the mass term, and
$$\eqalign{&T_{sc}= 2\sum_{n=1}^{\infty} {1\over n}
A^{\d}_{ij}(n)A_{ij}(n) \sum_{m=1}^{n-1} {(n+m)^2\over (n-m)^2 m} +\cr
& {1\over N}\sum_{n_i=1}^\infty \biggl\{
{\delta_{n_1+n_2, n_3+n_4}\over \sqrt{n_1 n_2 n_3 n_4}}
\left[ {(n_{2}-n_{1})(n_{4}-n_{3}) \over (n_{1} + n_{2})^2 } -
{(n_{3} + n_{1})(n_{4} + n_{2})
\over (n_{4}-n_{2})^2 } \right]
A_{kj}^{\d}(n_{3})A_{ji}^{\d}(n_{4})A_{kl}(n_{1})A_{li}(n_{2})  \cr
& + {\delta_{n_1+n_2+n_3,n_4}\over \sqrt{n_1 n_2 n_3 n_4}}
\left[ {(n_{1}+n_{4})(n_{3}-n_{2}) \over (n_{3}+n_{2})^2 } +
{(n_{3}+n_{4})(n_{1}-n_{2}) \over (n_{1}+n_{2})^2 } \right]\cr
&\bigl(A_{kj}^{\d}(n_{1})A_{jl}^{\d}(n_{2})
A_{li}^{\d}(n_{3})A_{ki}(n_{4}) + A_{kj}^{\d}(n_{4})A_{kl}(n_{1})A_{li}(n_{2})
A_{ij}(n_{3})\bigr) \biggr\} \cr
}\eqn\dmess$$
is the term generated by the gauge current-current interaction.
The cases where the denominator is zero are understood to be excluded
from the sum; this follows from the discretized analogue of the
principal value prescription. The only
dimensionless parameter in the problem is $y={4\pi m^2\over g^2 N}$, while
$g^2N$ sets the scale of the ``string tension'' . Large $y$ corresponds to
weak coupling, while $y\to 0$ is the strong coupling limit.

For the fermionic matter, we find
$$2P^+P^{-} = {g^2 N\over \pi}~K\left(x V_f +T_f\right)
\ ,\eqn\fham$$
where $x={\pi m^2\over g^2 N}$ is
the dimensionless parameter. Here the mass term $V_f$ is given by
eq. \pot\ with the $A$'s replaced by the $B$'s. The term generated
by the gauge interaction is now
$$\eqalign{&T_f= 2\sum_{n=1}^{\infty}
B^{\d}_{ij}(n)B_{ij}(n) \sum_{m=1}^{n-1} {1\over (n-m)^2 } +\cr
& {1\over N}\sum_{n_i=1}^\infty \biggl\{
\delta_{n_1+n_2, n_3+n_4}
\left[ {1\over (n_{4}-n_{2})^2 }
-{1\over (n_{1} + n_{2})^2 } \right]
B_{kj}^{\d}(n_{3})B_{ji}^{\d}(n_{4})B_{kl}(n_{1})B_{li}(n_{2})  \cr
& + \delta_{n_1+n_2+n_3,n_4}
\left[{1\over (n_{3}+n_{2})^2 } -
{1\over (n_{1}+n_{2})^2 } \right]\cr
&\bigl(
B_{kj}^{\d}(n_{4})B_{kl}(n_{1})B_{li}(n_{2}) B_{ij}(n_{3})
-B_{kj}^{\d}(n_{1})B_{jl}^{\d}(n_{2}) B_{li}^{\d}(n_{3})B_{ki}(n_{4})
\bigr) \biggr\} \cr
}\eqn\fmess$$

To illustrate the calculation of the (mass)$^2$
matrix with a simple example, consider the theory with the adjoint
fermions. We will set $K=6$ and study the even-bit (boson) and the
odd-bit (fermion) sectors separately: working to leading order
in $1/N$ the hamiltonian preserves the number of bits modulo 2.
For the even-bit sector the normalized states are
$$ \eqalign{
&|1>=
{1\over N^2}\Tr [B^\dagger(3) B^\dagger (1) B^\dagger(1)B^\dagger(1)]
|0>\ ,\qquad\quad
|2>={1\over N^2}\Tr [B^\dagger (2)B^\dagger (2)
 B^\dagger(1)B^\dagger(1)] |0>\ ,\cr
&|3>=
{1\over N^2\sqrt 2}\Tr [B^\dagger(2) B^\dagger (1) B^\dagger(2)B^\dagger(1)]
|0>\ ,\qquad\qquad
|4>={1\over N}\Tr [B^\dagger (5)B^\dagger (1) ] |0>\ ,\cr
&|5>={1\over N}\Tr [B^\dagger (4)B^\dagger (2)] |0>\ .
\cr }\eqn\eq$$
Our calculation gives
$$ xV_f+T_f=\left (\matrix { {10x\over 3} + {7\over 2} & 0 & 0 & 0 & 0 \cr
0 & 3x+ {653\over 144} & 0 & -{7\over 72} & {5\over 18} \cr
0 & 0 &  3x+{40\over 9} & 0 & 0 \cr
0 & -{7\over 72} & 0 & {6x\over 5}+{107\over 36} & -{16\over 9} \cr
0 & {5\over 18} & 0 & -{16\over 9} & {3x\over 4} + {47\over 9} } \right )
\ . \eqn\eq$$

Now for the odd-bit sector the normalized states are
$$ \eqalign{
&|1>=
{1\over N^{5/2}}
\Tr [B^\dagger(2) B^\dagger (1) B^\dagger(1)B^\dagger(1)B^\dagger(1)]
|0>\ ,\cr
&|2>={1\over N^{3/2}}\Tr [B^\dagger (4) B^\dagger(1)B^\dagger(1)] |0>\ ,
\qquad\qquad |3>=
{1\over N^{3/2}}\Tr [B^\dagger (3) B^\dagger(2)B^\dagger(1)]
|0>\ ,\cr
&|4>= {1\over N^{3/2}}\Tr [B^\dagger (3) B^\dagger(1)B^\dagger(2)]|0>\ ,
\qquad\qquad |5>=
{1\over N^{3/2}\sqrt 3}\Tr [B^\dagger (2) B^\dagger(2)B^\dagger(2)]|0>\ .
\cr }\eqn\eq$$
Here we find
$$ xV_f+T_f=\left (\matrix { {9x\over 2}+{43\over 36} & 0 & 0 & 0 & 0 \cr
0 & {9x\over 4}+{291\over 100} & -{117\over 100} & -{117\over 100} & 0 \cr
0 & -{117\over 100} & {11x\over 6}+{16969\over 3600} & -{7331\over 3600} &
-{15\sqrt{3}\over 16} \cr
0 & -{117\over 100} & -{7331\over 3600} & {11x\over 6} + {16969\over 3600} &
-{15\sqrt{3}\over 16} \cr
0 & 0 & -{15\sqrt{3}\over 16} & -{15\sqrt{3}\over 16} & {3x\over 2} +{99 \over
16} } \right ) \eqn\mat$$

For larger values of $K$ we used {\it Mathematica}
to generate the states, calculate and diagonalize
the (mass)$^2$ matrices for a range of values of $x,y$ and
$K$. At the values of $K$ up to which we were able to work,
we found very
poor convergence of the spectrum for the scalar matter.
The convergence improves
as $y$ increases toward  weak coupling,
since the $y \to \infty$ limit yields the free
particle spectrum, but no reliable data were obtainable in the
strong coupling region.
To illustrate this we plot in Figure~1 the
the (mass)$^2$ at $y=0$ for a low-lying state of scalar quanta in
the even-bit sector as
a function of $K$. In this and subsequent convergence plots the full matrix
was used up to $K=12$. In order to obtain some idea of
how convergence continues, we used the finding that low-lying
mass eigenstates are composed almost entirely of strings of some given length
(number of bits) to extrapolate
to higher $K$ by diagonalizing in the sub-space of bits of that length.
  For the state in Figure~1 we continued beyond $K=12$
by diagonalizing in the subspace of 2-bit strings, since for $K=12$
this state has probability of only
$\sim 10^{-6}$ of containing other than 2-bits. This is a
judicious choice since the 2-bit
space of states is especially simple and we could diagonalise up to $K\sim400$,
at which point the mass finally began to converge. Such a cutoff on the full
problem is way beyond the bounds of attainability.
Although quantitative conclusions are difficult to obtain in the strong
coupling region, we can say with some confidence that there are no
massless states in the $y\to 0$ limit. If we track
each state as a function of
the cut-off $K$, we find the mass monotonically increasing
with $K$.

The masses of the bound states of fermion quanta converge
extremely  well by comparison. Figures~2 and 3 show the full spectrum
of eigenvalues at $K=9$ as a function of $x$ for the even-bit  (bosonic)
and odd-bit (fermionic) sectors. (Close inspection reveals significant
bending of the paths.) The model still seems to be interacting fairly
strongly past $x=1$. Again there are many states of exceptional purity of
length. Figures~4 and 5 illustrate the convergence for the  lowest
eigenstates at
 $x=0$ and $x=0.8$ including both
sectors, extrapolating past $K=12$ by identifying the dominant length
component of the eigenfunction and diagonalizing in the subspace of states
of that length. The reader can identify these levels on Figures~2 and 3.
We conclude  that the latter are a fairly accurate reflection of the
very low-lying spectrum. One should keep in mind though, that for a cut-off
$K$ the results for states of length close to $K$ are dominated
by lattice artifacts, while states longer than $K$ do not appear at all.
Indeed for $K=12$ the masses of the
smallest length states are already close to converging, while as
the length increases the calculation gets less reliable. As $K$ grows,
more and more states are included in the scheme, and more masses are
obtained reliably.

To better quantify the  length-purity of these string eigenfunctions we plot
in Figure~6(i)  the low-lying spectrum in the
fermionic sector at $K=13$ and $x=0$
vs. average length of the bound states. It is tempting to identify
``Regge''-like trajectories in this diagram; horizontally we have excitations
of the length degree of freedom while vertically we have glue excitations.
Note, in particular, the rising line that can be drawn through the
purest states, whose average lengths are very close to $3, 5, 7, 9, ...$
This can be thought of as a ``trajectory'' connecting the length
excitations of the ground state. This trajectory appears to repeat for
the excited states, and in the continuum limit we may see an infinite
number of such trajectories.
In Figure~6(ii) we repeat the plots for $x=0.8$. This demonstrates
the effect of $x$ on the relation between the characteristic energies
for the glue and length excitations.

\bigskip
\centerline{\caps 5. Discussion}
\bigskip

Our discretized \lc\ analysis of the large-$N$ \td\ gauge
theory coupled to adjoint matter can certainly be improved by extending
the calculation to higher values of the cut-off $K$,
which requires a more efficient  computer program. We will however,
attempt to draw
some physical conclusions at this stage.
\item
{1.} For all values of $g/m$ the spectrum of (mass)$^2$ is positive and
appears to be discrete. There is no sign of a phase transition
for any value of the dimensionless parameter. This should be contrasted
with the behavior of the scalar matrix model with only global
$SU(N)$ symmetry. There we found evidence [\US] that the theory is tachyonic,
and that there is a phase transition, which
is associated with divergence of the sum of planar diagrams.
\item
{2.} Keeping $g$ fixed and sending $m$ to zero, we find no massless
states either for the scalar or for the fermion matter. It follows
that if $m$ is kept fixed while the gauge coupling $g$ is sent to
infinity, then the entire spectrum is pushed to infinite mass
and the theory becomes trivial. This behavior differs from
what is found in \td\ QCD with fundamental fermions, where there is a
massless state for $m=0$ [\qcd].
\item
{3.} The low-lying states are suprisingly close to being eigenstates
of the number of bits. For the fermionic matter this is certain to
be true in the continuum limit, since the convergence rate is good,
while for the scalar matter this property may also hold for
$K\to\infty$. This result could hardly be expected because
the matrix elements that change the number of bits are not small.
A qualitative picture that seems to emerge is that the low-lying
excitations can be roughly divided into two types:
pure glue excitations that do not alter the number of bits, and
excitations of the ``number of bits'' degree of freedom. The relation
between the two characteristic energy scales is, of course, a function
of $m/g$ (see Figures~6).
This separation is only approximate and breaks down for high
enough mass. The structure of states is therefore quite intricate:
they can be roughly separated into ``Regge trajectories'', which
group the glue excitations of the approximately constant ``bit state''.
This structure certainly requires further analysis. It is clearly
more complex  than in QCD with fundamental fermions,
where all the states consist of a quark and an antiquark so that
there is only one Regge trajectory [\qcd]. Coupling to the adjoint matter
gives rise to an additional ``number of bits'' degree of freedom,
which allows many new possibilities.
\item
{4.} The rate of convergence of the discretized \lc\ method is much better
for the fermions than for the scalars. This rate appears to be quite
sensitive to which theory is under study. Poor convergence in discretised
\lc\ quantisation, especially at small mass, is often associated with the
fact that low momentum quanta are not accurately accounted for at a given
$K$.\foot{At strictly zero mass the zero momentum quanta are allowed and
can even alter the vacuum from the naive Fock one [\Brod].}
Perhaps the models with
the adjoint matter will be of use as a testing ground for the
power of the method in higher-dimensional gauge theories.
It is also of great physical interest to extend our studies of the
longitudinal dynamics to gauge theories in dimensions greater than
two, where it becomes intertwined with transverse dynamics.
Theories of this type, with latticized transverse dimensions,
were studied in ref. [\BPR-\DM], and it is desirable to extend this analysis.

The models that we have analyzed can be regarded as close string
theories in the sense that they possess a \lc\ string description.
However, the absence of a phase transition makes it unlikely that
these theories have a continuum string formulation. We should,
in this case, look for gauged matrix models which do have critical
behavior. As we suggested in [\US], a good candidate model is
(in euclidean space)
$$ S_{gauged}=\int d^2 x \Tr \left ({1\over 4g^2}F_{\alpha\beta}^2+
\half (\partial_\alpha \phi+i[A_\alpha, \phi])^2+\half m^2 \phi^2+
{1\over 4}{\lambda\over N} \phi^4 \right )
\eqn\nac$$
which has a scalar self-interaction in addition to the gauge
field. Now there are two dimensionless parameters,
$g/m$ and $\lambda/m^2$. Criticality may occur for $\lambda<0$,
where the  number of scalar propagators in a typical
planar graph can begin to diverge. We have studied the model of
eq. \nac\ for $g=0$ and found results similar to those in the
$M^3$ model [\US], suggesting the possibility of a phase
transition at $\lambda=\lambda_c<0$.
Although the $g=0$ theory has problems due to the non-singlet states,
they can be eliminated by turning on arbitrary $g$, however small.
The effect of this modification on the singlet states vanishes in
the $g\to 0$ limit. Thus, eq. \nac\ with $g=0+$ appears to define a
model with a critical point where the string theory has an extra
continuous dimension. Its origin is that the ``number of bits'' degree
of freedom is tuned to criticality and becomes a massless field
on the string. Unfortunately, this string theory appears
to be tachyonic. Could it be that there is an entire critical line,
extending through non-zero values of $g$,  that in some
range passes through the non-tachyonic region of the $g-\lambda$ plane?
If we do find a
string theory
of a new type, it would be extremely interesting to find any
world sheet formulation.

A nice feature of the model with the fermion matter is that it gives
rise to strings that can be either space-time bosons or fermions.
An obvious extension is to formulate the supersymmetric matrix theory
in 1+1 dimensions, generalising the $c=1$ model of
 Marinari and Parisi [\mar].
This question and the other issues mentioned above are under investigation.

\ack
We thank P.Ginsparg, D.Gross and A. Polyakov for useful discussions.
I. R. K. is supported in part by
DOE grant DE-AC02-76WRO3072,
NSF Presidential Young Investigator Award PHY-9157482,
James S. McDonnell Foundation grant No. 91-48, and
an A. P. Sloan Foundation Research Fellowship. S. D. is supported by
S.E.R.C.(U.K.) post-doctoral fellowship RFO/B/91/9033.
\endpage

\centerline{FIGURE CAPTIONS}
\bigskip

\item
{\bf Fig.1.} A low-lying energy-level in the even-bit sector of the scalar
matter theory as a function of $K$. For $K \geq 13$ extrapolation is performed
by diagonalising in the 2-bit sector only.
\item
{\bf Fig.2.} The full spectrum of 29 states at $K=9$ for the odd-bit
(fermionic) sector of the fermionic matter theory as a function of $x$.
\item
{\bf Fig.3.} The full spectrum of 29 states at $K=9$ for the even-bit
(bosonic) sector of the fermionic matter theory as a function of $x$.
\item
{\bf Fig.4.} The lowest few energy-levels for fermionic matter (both even
(b) and odd (f) sectors) as a function of $K$ at $x=0$. From the bottom,
the sequence of statistics is $\{{\rm f,f,b,f,b,} \ldots\}, $
with extrapolation
for $K\geq 13$ using sectors of $\{3,5,2,7,4,\ldots \}$ bits.
\item
{\bf Fig.5.} The lowest few energy-levels for fermionic matter (both even
(b) and odd (f) sectors) as a function of $K$ at $x=0.8$. From the bottom,
the sequence of statistics is $\{{\rm f,b,f,f,b,} \ldots\}$,
with extrapolation
for $K\geq 13$ using sectors of $\{3,2,5,3,4,\ldots \}$ bits.
\item
{\bf Figs.6.} The lowest 15 energy-levels plotted against their average
length (number of quanta) for the odd-bit sector of fermionic matter at
$K=13$ and (i) $x=0$ (ii) $x=0.8$.
\endpage

\refout

\bye